\documentclass[preprintnumbers,amsmath,amssymbm,preprint]{revtex4}
\usepackage{epsfig}
\usepackage{graphicx}
\usepackage{amssymb}

\begin{document}
\title{Onset of superradiant instabilities in the composed Kerr-black-hole-mirror
bomb}
\author{Shahar Hod}
\affiliation{The Ruppin Academic Center, Emeq Hefer 40250, Israel}
\affiliation{ } \affiliation{The Hadassah Institute, Jerusalem
91010, Israel}
\date{\today}

\begin{abstract}
\ \ \ It was first pointed out by Press and Teukolsky that a system
composed of a spinning Kerr black hole surrounded by a reflecting
mirror may develop instabilities. The physical mechanism responsible
for the development of these exponentially growing instabilities is
the superradiant amplification of bosonic fields confined between
the black hole and the mirror. A remarkable feature of this composed
black-hole-mirror-field system is the existence of a critical mirror
radius, $r^{\text{stat}}_{\text{m}}$, which supports {\it
stationary} (marginally-stable) field configurations. This critical
(`stationary') mirror radius marks the boundary between stable and
unstable black-hole-mirror-field configurations: composed systems
whose confining mirror is situated in the region
$r_{\text{m}}<r^{\text{stat}}_{\text{m}}$ are stable (that is, all
modes of the confined field decay in time), whereas composed systems
whose confining mirror is situated in the region
$r_{\text{m}}>r^{\text{stat}}_{\text{m}}$ are unstable (that is,
there are confined field modes which grow exponentially over time).
In the present paper we explore this critical (marginally-stable)
boundary between stable and explosive black-hole-mirror-field
configurations. It is shown that the innermost ({\it smallest})
radius of the confining mirror which allows the extraction of
rotational energy from a spinning Kerr black hole approaches the
black-hole horizon radius in the extremal limit of rapidly-rotating
black holes. We find, in particular, that this critical mirror
radius (which marks the onset of superradiant instabilities in the
composed system) scales linearly with the black-hole temperature.

\end{abstract}
\bigskip
\maketitle


\section{Introduction}

One of the most intriguing phenomenon in black-hole physics is the
superradiant scattering of bosonic fields by spinning black holes:
it was first pointed out by Zel'dovich \cite{Zel} that a co-rotating
integer-spin wave field of frequency $\omega$ interacting with a
spinning Kerr black hole can be amplified (gain energy) if the
composed system is in the superradiant regime \cite{Noteun}
\begin{equation}\label{Eq1}
\omega<\omega_{\text{c}}\equiv m\Omega_{\text{H}}\  .
\end{equation}
Here $m$ is the azimuthal harmonic index of the incident wave field,
and
\begin{equation}\label{Eq2}
\Omega_{\text{H}}={{a}\over{r^2_++a^2}}\
\end{equation}
is the angular velocity of the spinning Kerr black hole
\cite{Notea}. The amplification of the scattered bosonic fields in
the superradiant regime (\ref{Eq1}) is accompanied by a decrease in
the rotational energy and angular momentum of the central spinning
black hole \cite{NoteRN,Bekch,Dego,Hodch,Li}.

Soon after Zel'dovich's discovery \cite{Zel} of the superradiance
phenomenon in black-hole physics, it was realized by Press and
Teukolsky \cite{PressTeu2} that the coupled black-hole-field system
may develop exponentially growing instabilities. This unstable
system, known as the {\it black-hole bomb}, is composed of three
ingredients \cite{PressTeu2}: (1) a spinning black hole whose
rotational energy serves as the energy source of the composed
system, (2) a co-rotating bosonic cloud which orbits the central
black hole and interacts with it to extract its rotational energy,
and (3) a reflecting mirror which surrounds the central black hole
and prevents the amplified bosonic field from radiating its energy
to infinity \cite{Notemas,Noteads}.

A remarkable feature of this composed physical system is the
existence, for each given set $(\bar a,l,m)$ of the black hole and
field parameters \cite{Notelm}, of a critical ({\it minimum}) mirror
radius which marks the boundary between stable and unstable
black-hole-mirror-field configurations. The critical (`stationary'
\cite{Noteccrit}) mirror radius, $r^{\text{stat}}_{\text{m}}(\bar
a,l,m)$, corresponds to {\it stationary} confined field
configurations which are characterized by the critical superradiant
frequency (\ref{Eq1}).

The composed black-hole-mirror-scalar-field system was studied in
the important work by Cardoso et. al. \cite{CarDias,Notehs}. In
particular, it was shown in \cite{CarDias} that, for a given set of
parameters $(\bar a,l,m)$, black-hole-mirror-field systems whose
mirror radii lie in the regime
$r_{\text{m}}<r^{\text{stat}}_{\text{m}}(\bar a,l,m)$ are stable
(the confined scalar mode decays in time) whereas
black-hole-mirror-field systems whose mirror radii lie in the regime
$r_{\text{m}}>r^{\text{stat}}_{\text{m}}(\bar a,l,m)$ are unstable
(the confined scalar mode grows exponentially over time).

The numerical results presented in \cite{CarDias} indicate that the
critical (`stationary' \cite{Noteccrit}) mirror radius,
$r^{\text{stat}}_{\text{m}}(\bar a,l,m)$, has the following three
important features:
\begin{itemize}
\item{For a confined field mode of given harmonic indexes $(l,m)$ [see Eq. (\ref{Eq10}) below], the
stationary mirror radius is a decreasing function of the black-hole
angular momentum $\bar a$. In other words, rapidly-rotating Kerr
black holes are characterized by stationary mirror radii which are
smaller (closer to the black-hole horizon) than the corresponding
stationary mirror radii of slowly-rotating black holes.}
\item{For a given value of the spheroidal harmonic index $l$ of the
confined field, the stationary mirror radius decreases with
increasing values of the azimuthal harmonic index $m$. The
equatorial $m=l$ mode is therefore characterized by the innermost
(smallest) stationary mirror radius among confined field modes which
share the same spheroidal harmonic index $l$.}
\item{For confined equatorial $m=l$ modes, the stationary mirror radius decreases
with increasing values of the spheroidal harmonic index $l$.}
\end{itemize}

From these characteristics of the stationary mirror radius, one
concludes that the asymptotic radius
\begin{equation}\label{Eq3}
r^*_{\text{m}}\equiv r^{\text{stat}}_{\text{m}}(\bar a\to
1,l=m\to\infty)\
\end{equation}
provides the innermost location of the confining mirror. The
physical significance of this critical mirror radius,
$r^*_{\text{m}}$, lies in the fact that this is the innermost ({\it
smallest}) radius of the confining mirror which allows the
extraction of rotational energy from spinning Kerr black holes.
Below we shall explore the physical properties of this critical
mirror radius.

Cardoso et. al. \cite{CarDias} also provided an analytic treatment
of the black-hole-mirror-field system in the {\it small} frequency
regime $a\omega\ll1$. Substituting the critical (marginal)
superradiant frequency $\omega_{\text{c}}=m\Omega_{\text{H}}$ [see
Eq. (\ref{Eq1})] into equation (30) of \cite{CarDias}, one finds
that the stationary mirror radius $r^{\text{stat}}_{\text{m}}(\bar
a,l,m)$ is given as a solution of the characteristic equation
\begin{equation}\label{Eq4}
J_{l+1/2}(m\Omega_{\text{H}}r^{\text{stat}}_{\text{m}})=0\  ,
\end{equation}
where $J_{l+1/2}(x)$ is the Bessel function. It should be emphasized
that the characteristic equation (\ref{Eq4}) for the `stationary'
(critical) radii of the mirror is valid in the regime $a\omega\ll1$
considered in \cite{CarDias}. Thus, the characteristic equation
(\ref{Eq4}) can only determine the stationary mirror radii of {\it
slowly}-rotating black holes in the regime $m\bar a\ll1$.

One of the goals of the present study is to obtain an analogous
characteristic equation for the stationary mirror radii of {\it
rapidly}-rotating ($\bar a \simeq 1$) black holes. As discussed
above, the {\it numerical} results presented in \cite{CarDias}
indicate that these near-extremal black holes are characterized by
stationary mirror radii which are {\it closer} to the black-hole
horizon than the corresponding stationary mirror radii (\ref{Eq4})
of slowly-rotating black holes. Below we shall confirm this
expectation {\it analytically}.

In addition, as we shall show below, our characteristic equation for
the critical (`stationary' \cite{Noteccrit}) mirror radii of
rapidly-rotating black holes [see Eq. (\ref{Eq21}) below] is valid
for arbitrarily large values of the harmonic indexes $(l,m)$ of the
confined field \cite{Notelim}. This fact will allow us to address
the following interesting question regarding the nature of this
critical mirror radius: what is the {\it asymptotic} behavior of the
stationary mirror radius in the eikonal $l\gg1$ limit?

As discussed above, the asymptotic mirror radius,
$r^*_{\text{m}}\equiv r^{\text{stat}}_{\text{m}}(\bar a\to
1,l=m\to\infty)$, corresponds to the {\it innermost} location of the
confining mirror which allows the extraction of rotational energy
from spinning Kerr black holes. In addition, for generic confined
field configurations \cite{Noteclo}, this critical mirror radius
marks the boundary between stable and explosive
black-hole-mirror-field configurations. One of the goals of the
present study is to determine this fundamental (asymptotic) mirror
radius $r^*_{\text{m}}$.

\section{Description of the system}

The explored physical system is composed of a spinning Kerr black
hole of mass $M$ and angular momentum $Ma$ linearly coupled to a
massless scalar field $\Psi$. In the Boyer-Lindquist coordinate
system $(t,r,\theta,\phi)$ the black-hole spacetime geometry is
described by the line-element \cite{Chan,Kerr}
\begin{eqnarray}\label{Eq5}
ds^2=-\Big(1-{{2Mr}\over{\rho^2}}\Big)dt^2-{{4Mar\sin^2\theta}\over{\rho^2}}dt
d\phi+{{\rho^2}\over{\Delta}}dr^2
+\rho^2d\theta^2+\Big(r^2+a^2+{{2Ma^2r\sin^2\theta}\over{\rho^2}}\Big)\sin^2\theta
d\phi^2,
\end{eqnarray}
where $\Delta\equiv r^2-2Mr+a^2$ and $\rho^2\equiv
r^2+a^2\cos^2\theta$. The zeroes of $\Delta$ determine the radii of
the black-hole (event and inner) horizons:
\begin{equation}\label{Eq6}
r_{\pm}=M\pm(M^2-a^2)^{1/2}\  .
\end{equation}

As discussed above, the critical (`stationary' \cite{Noteccrit})
mirror radius which characterizes the composed
black-hole-mirror-field system is a decreasing function of the
black-hole rotation parameter $\bar a$. Thus, rapidly-rotating black
holes are expected to be characterized by the smallest ({\it
innermost}) stationary mirror radii. In the present study we shall
analyze the physical properties of the black-hole-mirror-field
system in this physically interesting regime of rapidly-rotating
Kerr black holes with
\begin{equation}\label{Eq7}
{\bar a}\simeq 1\  .
\end{equation}

The dynamics of the scalar field $\Psi$ in the curved geometry is
determined by the Klein-Gordon wave equation
\begin{equation}\label{Eq8}
\nabla^a \nabla_a \Psi=0\  ,
\end{equation}
which, in the rotating Kerr spacetime (\ref{Eq5}) becomes
\cite{Teu,Noteknt}
\begin{eqnarray}\label{Eq9}
\Big[{{(r^2+a^2)^2}\over{\Delta}}-a^2\sin^2\theta\Big]
{{\partial^2\Psi}\over{\partial
t^2}}+{{4Mar}\over{\Delta}}{{\partial^2\Psi}\over{\partial t
\partial\phi}}+\Big({{a^2}\over{\Delta}}
-{{1}\over{\sin^2\theta}}\Big){{\partial^2\Psi}\over{\partial\phi^2}}
-\Delta{{\partial}\over{\partial
r}}\Big(\Delta{{\partial\Psi}\over{\partial r}}\Big)\nonumber
\\ -{{1}\over{\sin\theta}}{{\partial}\over{\partial\theta}}\Big(\sin\theta
{{\partial\Psi}\over{\partial\theta}}\Big)=0\  .
\end{eqnarray}

It is convenient to decompose the scalar field $\Psi$ in the form
\begin{equation}\label{Eq10}
\Psi=\sum_{l,m}e^{im\phi}{S_{lm}}(\theta;a\omega){R_{lm}}(r;a,\omega)e^{-i\omega
t}\ ,
\end{equation}
in which case one finds \cite{Teu} that the radial $R_{lm}$ and
angular $S_{lm}$ wave functions are determined by two coupled
ordinary  differential equations of the confluent Heun type
\cite{TeuPre2,Heun,Flam,Fiz1}, see Eqs. (\ref{Eq11}) and
(\ref{Eq13}) below.

It is worth emphasizing that the sign of $\Im\omega$ in (\ref{Eq10})
reflects the stability/instability properties of the confined field
mode: stable modes (modes which decay exponentially in time) are
characterized by $\Im\omega<0$, whereas unstable modes (modes which
grow exponentially over time) are characterized by $\Im\omega>0$.
Stationary (marginally-stable) field modes are characterized by
$\Im\omega=0$.

The angular functions ${S_{lm}}(\theta;a\omega)$ are known as the
spheroidal harmonic functions. These functions are solutions of the
angular differential equation \cite{TeuPre2,Heun,Flam,Hodop}
\begin{equation}\label{Eq11}
{1\over {\sin\theta}}{d \over {d\theta}}\Big(\sin\theta {{d
S_{lm}}\over{d\theta}}\Big)+\Big(a^2\omega^2\cos^2\theta-{{m^2}\over{\sin^2\theta}}+A_{lm}\Big)S_{lm}=0\
\end{equation}
in the interval $\theta\in [0,\pi]$. Regular eigenfunctions
\cite{Noteregu} exist for a discrete set $\{A_{lm}(a\omega)\}$ of
angular eigenvalues which are labeled by the two integers $m$ and
$l\geq|m|$. These angular eigenvalues can be expanded in the form
\cite{TeuPre1,BerAlm,Abram}
\begin{equation}\label{Eq12}
{A_{lm}}(a\omega)=l(l+1)+\sum_{k=1}^{\infty}c_k(a\omega)^k\  ,
\end{equation}
where the expansion coefficients $\{c_k(l,m)\}$ are given in
\cite{TeuPre1,BerAlm,Abram}.

The radial functions ${R_{lm}}(r;a,\omega)$ satisfy the ordinary
differential equation \cite{TeuPre2}
\begin{equation}\label{Eq13}
\Delta{{d} \over{dr}}\Big(\Delta{{dR_{lm}}\over{dr}}\Big)+\Big[K^2
-\Delta(a^2\omega^2-2ma\omega+A_{lm})\Big]R_{lm}=0\ ,
\end{equation}
where $K\equiv(r^2+a^2)\omega-ma$. Note that the angular
differential equation (\ref{Eq11}) and radial differential equation
(\ref{Eq13}) are coupled by the angular eigenvalues
$\{{A_{lm}}(a\omega)\}$. (We shall henceforth omit the harmonic
indexes $l$ and $m$ for brevity.)

We shall be interested in solutions of the radial wave equation
(\ref{Eq13}) with the physical requirement (boundary condition) of
purely ingoing waves (as measured by a comoving observer) crossing
the black-hole horizon \cite{TeuPre2}. As shown in \cite{TeuPre2},
this boundary condition corresponds to the behavior
\begin{equation}\label{Eq14}
R \sim e^{-i (\omega-m\Omega_{\text{H}})y}\ \ \ \text{as}\ \ \ r\to
r_+\ \ (y\to -\infty)\
\end{equation}
of the radial eigenfunction in the vicinity of the black-hole
horizon, where the ``tortoise" radial coordinate $y$ is defined by
$dy=[(r^2+a^2)/\Delta]dr$. In addition, following
\cite{CarDias,Hodch} we shall assume that the scalar field vanishes
at the location $r_{\text{m}}$ of the confining mirror:
\begin{equation}\label{Eq15}
R(r=r_{\text{m}})=0\  .
\end{equation}

For the analysis of the radial Teukolsky equation (\ref{Eq13}), it
is convenient to define new dimensionless variables
\begin{equation}\label{Eq16}
x\equiv {{r-r_+}\over {r_+}}\ \ ;\ \ \tau\equiv 8\pi
MT_{\text{BH}}={{r_+-r_-}\over {r_+}}\ \ ;\ \ k=2\omega r_+\ \ ; \ \
\varpi\equiv {{\omega-m\Omega_{\text{H}}}\over{2\pi T_{\text{BH}}}}\
,
\end{equation}
in terms of which Eq. (\ref{Eq13}) becomes
\begin{equation}\label{Eq17}
x(x+\tau){{d^2R}\over{dx^2}}+(2x+\tau){{dR}\over{dx}}+VR=0\  ,
\end{equation}
where $V\equiv K^2/r^2_+x(x+\tau)-(a^2\omega^2-2ma\omega+A)$ and
$K=r^2_+\omega x^2+r_+kx+r_+\varpi\tau/2$.

\section{Marginally stable black-hole-mirror-field configurations}

We shall now analyze the {\it stationary} ($\Im\omega=0$) resonances
of the composed system. As discussed above, these stationary
(marginally-stable) black-hole-mirror-field configurations are
characterized by the critical frequency (\ref{Eq1}) for superradiant
scattering in the black-hole spacetime. In particular, in this
section we shall find the {\it discrete} set of mirror radii,
$\{r^{\text{stat}}_{\text{m}}(\bar a,l,m;n)\}$, which support
stationary confined field configurations. (Here $n=1,2,3,...$ is the
resonance parameter).

We consider a rapidly-rotating Kerr black hole \cite{Noterap}
surrounded by a reflecting mirror which is placed in the vicinity of
the black-hole horizon. In particular, we shall assume the following
inequalities:
\begin{equation}\label{Eq18}
\tau\ll1\ \ \ \text{and}\ \ \ x_{\text{m}}\ll1\  .
\end{equation}

In the near-horizon region $x\ll1$ the radial equation is given by
(\ref{Eq17}) with $V\to
V_{\text{near}}\equiv-(a^2\omega^2-2ma\omega+A)+(kx+\varpi\tau/2)^2/x(x+\tau)$.
The physical radial solution obeying the ingoing boundary condition
(\ref{Eq14}) at the black-hole horizon with the critical
(marginally-stable) superradiant frequency $\varpi=0$ is given by
\cite{TeuPre2,Abram}
\begin{equation}\label{Eq19}
R(x)=\Big({x\over \tau}+1\Big)^{-ik}{_2F_1}({1\over
2}-ik+i\delta,{1\over 2}-ik-i\delta;1;-x/\tau)\  ,
\end{equation}
where $_2F_1(a,b;c;z)$ is the hypergeometric function \cite{Abram},
and
\begin{equation}\label{Eq20}
\delta^2\equiv -a^2\omega^2+2ma\omega-A+k^2-{1\over 4}\ .
\end{equation}
We shall henceforth consider the case of real $\delta$
\cite{Notedeq,Notedel}. The mirror-like boundary condition
$R(x=x^{\text{stat}}_{\text{m}})=0$ [see Eq. (\ref{Eq15})] now reads
\begin{equation}\label{Eq21}
{_2F_1}(1/2-ik+i\delta,1/2-ik-i\delta;1;-x^{\text{stat}}_{\text{m}}/\tau)=0\
.
\end{equation}

It is worth emphasizing that, the newly derived resonance condition
(\ref{Eq21}) for the critical (`stationary' \cite{Noteccrit}) mirror
radii of the system is valid for confining mirrors which are placed
in the near-horizon region $x_{\text{m}}\ll1$. [Below we shall see
that this near-horizon condition implies that the stationary mirror
radii obtained from (\ref{Eq21}) are valid in the regime of {\it
rapidly}-rotating black holes with $\tau\ll1$]. On the other hand,
the resonance condition (\ref{Eq4}) \cite{CarDias} for the
stationary mirror radii is valid in the complementary regime
$x_{\text{m}}\gg1$, or equivalently in the regime of {\it
slowly}-rotating black holes with $m\bar a\ll1$.

One important conclusion which can immediately be drawn from the
resonance condition (\ref{Eq21}) is the fact that, for
rapidly-rotating black holes, the critical mirror radius scales
linearly with the black-hole temperature \cite{Notelin}:
\begin{equation}\label{Eq22}
x^{\text{stat}}_{\text{m}}\propto\tau\  .
\end{equation}
This implies that the stationary mirror radius
$x^{\text{stat}}_{\text{m}}$ is a decreasing function of the
black-hole rotation parameter $\bar a$ (an increasing function of
the black-hole dimensionless temperature $\tau$).

As we shall now show, for small and moderate values of the field
harmonic indexes $(l,m)$, the stationary mirror radii
$x^{\text{stat}}_{\text{m}}(\bar a,l,m;n)$ are characterized by the
relation
\begin{equation}\label{Eq23}
{{x^{\text{stat}}_{\text{m}}}/{\tau}}\gg1\  ,
\end{equation}
in which case the resonance condition (\ref{Eq21}) can be solved
{\it analytically}. In the regime (\ref{Eq23}) one can use the
large-$z$ asymptotic behavior of the hypergeometric function
$_2F_1(a,b;c;z)$ \cite{Abram} to approximate the resonance condition
(\ref{Eq21}) by
\begin{equation}\label{Eq24}
{{\Gamma(2i\delta)}\over{\Gamma(1/2-ik+i\delta)
\Gamma(1/2+ik+i\delta)}}
\Big({{x^{\text{stat}}_{\text{m}}}\over{\tau}}\Big)^{i\delta}+{{\Gamma(-2i\delta)}\over{\Gamma(1/2-ik-i\delta)
\Gamma(1/2+ik-i\delta)}}
\Big({{x^{\text{stat}}_{\text{m}}}\over{\tau}}\Big)^{-i\delta}=0\  ,
\end{equation}
which yields
\begin{equation}\label{Eq25}
\Big({{x^{\text{stat}}_{\text{m}}}\over{\tau}}\Big)^{2i\delta}
=-{{\Gamma(-2i\delta)\Gamma(1/2-ik+i\delta)\Gamma(1/2+ik+i\delta)}\over{\Gamma(2i\delta)\Gamma(1/2-ik-i\delta)
\Gamma(1/2+ik-i\delta)}} .
\end{equation}

It is easy to verify that, for real values of the angular eigenvalue
$\delta$, the characteristic equation (\ref{Eq25}) corresponds to
real values of the stationary mirror radii \cite{Notehs}: taking the
logarithm of both sides of (\ref{Eq25}), one obtains the
characteristic equation \cite{Notem1}
\begin{equation}\label{Eq26}
2\delta\ln\Big({{x^{\text{stat}}_{\text{m}}}\over{\tau}}\Big)=i\ln\Big[{{\Gamma(2i\delta)}\over{\Gamma(-2i\delta)}}\Big]+
i\ln\Big[{{\Gamma(1/2-ik-i\delta)}\over{\Gamma(1/2+ik+i\delta)}}\Big]+
i\ln\Big[{{\Gamma(1/2+ik-i\delta)}\over{\Gamma(1/2-ik+i\delta)}}\Big]+\pi(2n-1)\
\end{equation}
for the discrete set $\{x^{\text{stat}}_{\text{m}}(\bar a,l,m;n)\}$
of stationary mirror radii, where $n=1,2,3,...$ is the resonance
parameter of the mode. Inspection of the resonance condition
(\ref{Eq26}) reveals the following facts: the first three terms on
the r.h.s of this equation are of the form $i[\ln(z)-\ln(\bar z)]$
and are therefore purely real numbers (see Eq. 6.1.23 of
\cite{Abram}). The fourth term on the r.h.s of (\ref{Eq26}) is
obviously a purely real number. This implies that the r.h.s of
(\ref{Eq26}) is a purely real number. One therefore concludes that,
for real values of the angular eigenvalue $\delta$, the
dimensionless mirror radii obtained from the resonance condition
(\ref{Eq26}) are purely real numbers. These stationary
\cite{Noteccrit} mirror radii are given by
\begin{equation}\label{Eq27}
x^{\text{stat}}_{\text{m}}(n)=\tau\times
\Big[{{\Gamma(-2i\delta)\Gamma(1/2+ik+i\delta)\Gamma(1/2-ik+i\delta)}\over
{\Gamma(2i\delta)\Gamma(1/2-ik-i\delta)\Gamma(1/2+ik-i\delta)}}\Big]^{1/2i\delta}e^{\pi(2n-1)/2\delta}\
.
\end{equation}

In Table \ref{Table1} we display the discrete radii of the
reflecting mirror corresponding to the stationary confined
equatorial $l=m=2$ mode as obtained from the analytical formula
(\ref{Eq27}) \cite{Noten0}. We also display the corresponding mirror
radii as obtained from a direct numerical solution \cite{Noteeas} of
the characteristic resonance equation (\ref{Eq21}). One finds a
remarkably good agreement between the exact mirror radii [as
obtained numerically from the resonance condition (\ref{Eq21})] and
the approximated mirror radii [as obtained from the analytical
formula (\ref{Eq27})].

\begin{table}[htbp]
\centering
\begin{tabular}{|c|c|c|c|c|c|}
\hline \text{Formula} & \ $x^{\text{stat}}_{\text{m}}(n=1)/\tau$\ \
& \ $x^{\text{stat}}_{\text{m}}(n=2)/\tau$\ \ & \
$x^{\text{stat}}_{\text{m}}(n=3)/\tau$\ \ &
\ $x^{\text{stat}}_{\text{m}}(n=4)/\tau$\ \ \\
\hline
\ {\text{Analytical}}\ [Eq. (\ref{Eq27})]\ \ &\ 94.06\ \ &\ 2613.20\ \ &\ 72603.50\ \ &\ $2017168.1$\\
\ {\text{Numerical}}\ [Eq. (\ref{Eq21})]\ \ &\ 91.79\ \ &\ 2610.95\ \ &\ 72601.28\ \ &\ $2017165.8$\\
\hline
\end{tabular}
\caption{Stationary resonances of the composed
black-hole-mirror-field system. We display the scaled mirror radii,
$x^{\text{stat}}_{\text{m}}(l=m=2;n)/\tau$, corresponding to the
stationary confined mode with $l=m=2$, as obtained from a direct
numerical solution of the characteristic resonance equation
(\ref{Eq21}). We also display the corresponding mirror radii as
obtained from the analytical formula (\ref{Eq27}). One finds a
remarkably good agreement between the exact (numerically computed)
mirror radii and the approximated (analytically calculated) mirror
radii.} \label{Table1}
\end{table}

In Table \ref{Table2} we display the discrete set of mirror radii,
$\{x^{\text{stat}}_{\text{m}}(l=m;n)/\tau\}$, corresponding to
composed black-hole-mirror configurations with stationary confined
equatorial ($l=m$) field modes \cite{Notemm,Notetau,Notedel2}. The
data presented in Table \ref{Table2} correspond to a direct
numerical solution \cite{Noteeas} of the characteristic resonance
condition (\ref{Eq21}). It is worth recalling that, for a confined
pure field (that is, a confined mode characterized by a given value
of the azimuthal harmonic index $m$), the smallest stationary mirror
radius, $x^{\text{stat}}_{\text{m}}(\bar a,m;n=1)$, marks the
boundary between stable and unstable black-hole-mirror-field
configurations for that particular pure mode.

From Table \ref{Table2} one learns that the stationary mirror
radius, $x^{\text{stat}}_{\text{m}}(\bar a,m;n=1)$, is a {\it
decreasing} function of the azimuthal harmonic index $m$. This is a
generic feature of the composed black-hole-mirror-field system: in
Table \ref{Table3} we display the values of the stationary mirror
radii in the asymptotic $m\gg 1$ regime \cite{Notef,Hoddel}. One
finds that the data presented in Table \ref{Table3} is described
extremely well by the simple asymptotic formula:
\begin{equation}\label{Eq28}
{{x^{\text{stat}}_{\text{m}}}(\bar a\to 1,
l=m\gg1)\over{\tau}}\simeq \alpha+{{\beta}\over{m}}+O(m^{-2})\ \ \
{\text{with}}\ \ \ \alpha\simeq 0.36 \ \ ; \ \ \beta\simeq 10.5\ .
\end{equation}

What we find most interesting is the fact that the coefficient
$\alpha$ in (\ref{Eq28}) has a finite asymptotic value. This fact
suggests that the composed Kerr-black-hole-mirror-field system is
characterized by a {\it finite} asymptotic limit of the critical
(`stationary') mirror radius:
\begin{equation}\label{Eq29}
{{x^*_{\text{m}}}\over{\tau}}\simeq 0.36\  ,
\end{equation}
where $x^*_{\text{m}}\equiv x^{\text{stat}}_{\text{m}}(\bar a\to
1,l=m\to\infty)$ [see Eq. (\ref{Eq3})].

This feature of the {\it spinning} black-hole-mirror-field bomb
should be contrasted with the corresponding case of the {\it
charged} black-hole-mirror-field bomb \cite{Dego,Hodch}. In the
charged case one finds \cite{Hodch} that the critical (`stationary')
mirror radius $x^{\text{stat}}_{\text{m}}$ can be made arbitrarily
small (that is, the mirror can be placed arbitrarily close to the
black-hole horizon) in the asymptotic \cite{Notecharge} $qQ\gg 1$
regime \cite{Hodch}:
\begin{equation}\label{Eq30}
{{x^{\text{stat}}_{\text{m}}(qQ\gg1)}\over{\tau}}=O(1/qQ)\to 0\  .
\end{equation}

\begin{table}[htbp]
\centering
\begin{tabular}{|c|c|c|c|c|c|c|}
\hline $\ l=m\ $ & $\delta$ & \
$x^{\text{stat}}_{\text{m}}(n=1)/\tau$\ \ & \
$x^{\text{stat}}_{\text{m}}(n=2)/\tau$\ \ & \
$x^{\text{stat}}_{\text{m}}(n=3)/\tau$\ \ &
\ $x^{\text{stat}}_{\text{m}}(n=4)/\tau$\ \ \\
\hline
\ 2\ \ &\ \ 0.945\ \ \ &\ 91.79\ \ &\ 2610.95\ \ &\ 72601.28\ \ &\ $2017165.8$\\
\ 3\ \ &\ \ 1.937\ \ \ &\ 10.99\ \ &\ 62.43\ \ &\ 322.69\ \ &\ 1640.27\\
\ 4\ \ &\ \ 2.849\ \ \ &\ 5.23\ \ &\ 18.85\ \ &\ 59.77\ \ &\ 183.00\\
\ 5\ \ &\ \ 3.739\ \ \ &\ 3.45\ \ &\ 9.95\ \ &\ 24.93\ \ &\ 59.59\\
\hline
\end{tabular}
\caption{Stationary resonances of the composed
black-hole-mirror-field system. We display the dimensionless mirror
radii, $x^{\text{stat}}_{\text{m}}(m;n)/\tau$, corresponding to
stationary confined modes. The data presented is for the equatorial
$l=m$ modes, the modes with the smallest (innermost) stationary
mirror radii. Also shown are the corresponding values of the angular
eigenvalues $\delta$ [see Eq. (\ref{Eq20})]. One finds that the
critical (stationary) radius of the mirror,
$x^{\text{stat}}_{\text{m}}(m;n)$, is a decreasing function of the
azimuthal harmonic index $m$.} \label{Table2}
\end{table}

\begin{table}[htbp]
\centering
\begin{tabular}{|c|c|c|c|c|c|c|c|c|}
\hline
\ \ $l=m$\ \ & \ 25\ \ & \ 50\ \ & \ 75\ \ & \ 100\ \ & \ 125\ \ & \ 150\ \ & \ 175\ \ & \ 200\ \ \\
\hline \ \ $x^{\text{stat}}_{\text{m}}(m;n=1)/\tau$\ \ \ & \ \ 0.75\
\ \ & \ \ 0.55\ \ \ & \ \ 0.49\ \ \ & \ \ 0.46\ \ \ & \ \ 0.44\ \ \
& \ \ 0.43\ \ \
& \ \ 0.42 \ \ \ & \ \ 0.41\ \ \ \\
\hline
\end{tabular}
\caption{Stationary resonances of the composed
black-hole-mirror-field system. We display the dimensionless mirror
radii, $x^{\text{stat}}_{\text{m}}(m;n=1)/\tau$, corresponding to
stationary confined modes in the asymptotic $l=m\gg 1$ regime. One
finds that the critical (stationary) radius of the mirror,
$x^{\text{stat}}_{\text{m}}(m;n=1)$, decreases monotonically to an
asymptotic {\it finite} value [see Eq. (\ref{Eq29})] in the eikonal
$l\gg1$ limit.} \label{Table3}
\end{table}

In order to support our findings, according to which the composed
Kerr-black-hole-mirror-field system is characterized by a {\it
finite} asymptotic value of the critical mirror radius, we shall
prove in the next section that composed systems whose mirror radii
lie in the regime $x_{\text{m}}/\tau\ll 1$ {\it cannot} support
stationary confined field configurations.

\section{No stationary confined field configurations in the regime $x_{\text{m}}/\tau\ll 1$}

We have seen that the (scaled) critical mirror radius,
$x^{\text{stat}}_{\text{m}}(m)/\tau$, decreases monotonically with
increasing values of the azimuthal harmonic index $m$. This fact
naturally gives rise to the following question: Can the ratio
$x^{\text{stat}}_{\text{m}}(m)/\tau$ be made arbitrarily small in
the asymptotic eikonal regime $m\to\infty$? One can also formulate
this question in a more practical form: Is it possible to extract
the black-hole rotational energy by placing a confining mirror
arbitrarily close to its horizon \cite{Noteprac}?

Our analysis in Sec. III provides compelling evidence that the
answer to the above questions is ``no". In particular, our results
suggest that the Kerr-black-hole-mirror-field system is
characterized by a {\it finite} asymptotic value of the critical
(`stationary' \cite{Noteccrit}) mirror radius:
$x^{\text{stat}}_{\text{m}}(m\gg1)/\tau=O(1)$ [see Eq.
(\ref{Eq29})]. In order to support this finding, we shall prove in
this section that composed black-hole-mirror systems whose mirror
radii lie in the regime $x_{\text{m}}/\tau\ll 1$ cannot support
stationary confined field configurations \cite{Noteminm}.

To that end, it proves useful to write the radial Teukolsky equation
in the form of a Schr\"odinger-like wave equation \cite{PressTeu3}:
\begin{equation}\label{Eq31}
{{d^2\psi}\over{dy^2}}+V\psi=0\  ,
\end{equation}
where $\psi=(r^2+a^2)^{1/2}R$ and the ``tortoise" radial coordinate
$y$ was defined in Sec. II. The potential $V$ in Eq. (\ref{Eq31}) is
given by \cite{PressTeu3}
\begin{equation}\label{Eq32}
V={{K^2-\Delta\lambda}\over{(r^2+a^2)^2}}-G^2-{{dG}\over{dy}}\  ,
\end{equation}
where
\begin{equation}\label{Eq33}
G={{r\Delta}\over{(r^2+a^2)^2}}\  ,
\end{equation}
and
\begin{equation}\label{Eq34}
\lambda=k^2-{1\over 4}-\delta^2\  .
\end{equation}
In the near-horizon region \cite{Notene}
\begin{equation}\label{Eq35}
x\ll\tau\  ,
\end{equation}
one finds \cite{Notey}
\begin{equation}\label{Eq36}
y\simeq
{{r^2_++a^2}\over{r_+-r_-}}\ln\Big({{r-r_+}\over{r_+-r_-}}\Big)={{2M}\over{\tau}}\ln\big({x\over\tau}\big)\
,
\end{equation}
which implies $\Delta\simeq (r_+-r_-)^2e^{\tau y/2M}$ and
\begin{equation}\label{Eq37}
V(y)\simeq (\omega-m\Omega_{\text{H}})^2-V_{\text{H}}e^{\tau y/2M}\
,
\end{equation}
where
\begin{equation}\label{Eq38}
V_{\text{H}}\equiv\Big({{\tau}\over{2M}}\Big)^2\Big(\lambda+{{\tau
r_+}\over{2M}}\Big)\ .
\end{equation}

In the eikonal regime, $l,m\gg1$, one can use the relation
\cite{Hoddel}
\begin{equation}\label{Eq39}
\delta=l\times \sqrt{-1+{15\over 8}\mu^2-{1\over 8}\mu^4}+O(1)\ \ \
; \ \ \ \mu\equiv {m \over l}\  ,
\end{equation}
in order to find [see Eq. (\ref{Eq34})]
\begin{equation}\label{Eq40}
\lambda=l^2\Big(1-{7\over 8}\mu^2+{1\over 8}\mu^4\Big)\  .
\end{equation}
From (\ref{Eq40}) one concludes that
\begin{equation}\label{Eq41}
\lambda>0\
\end{equation}
for all values of the dimensionless ratio $\mu$ (note that
$\mu\leq1$).

Taking cognizance of Eqs. (\ref{Eq31}), (\ref{Eq37}), and
(\ref{Eq38}), one obtains the radial wave equation
\begin{equation}\label{Eq42}
{{d^2\psi}\over{dy^2}}-V_{\text{H}}e^{\tau y/2M}\psi=0\
\end{equation}
for stationary field configurations (with
$\omega=m\Omega_{\text{H}}$). Defining
\begin{equation}\label{Eq43}
z={{\tau}\over{4M}}y\  ,
\end{equation}
the radial wave equation (\ref{Eq42}) can be written in the form
\begin{equation}\label{Eq44}
{{d^2\psi}\over{dz^2}}-4\Big(\lambda+{{\tau r_+}\over{2M}}\Big)
e^{2z}\psi=0\  .
\end{equation}
Using Eq. (9.1.54) of \cite{Abram}, one finds that the physical
solution to the radial equation (\ref{Eq44}) is described by the
Bessel function of the first kind:
\begin{equation}\label{Eq45}
\psi(z)=J_0\Big(2i\sqrt{\lambda+{{\tau r_+}\over{2M}}}e^z\Big)\  .
\end{equation}
This radial solution can also be written [see Eqs. (\ref{Eq36}) and
(\ref{Eq43})] in the form \cite{Notext}
\begin{equation}\label{Eq46}
\psi(x)=J_0\Big(2i\sqrt{\Big(\lambda+{{\tau
r_+}\over{2M}}\Big){{x}\over{\tau}}}\Big)\ .
\end{equation}

Taking cognizance of Eq. (\ref{Eq41}), one finds that the argument
of the Bessel function in (\ref{Eq46}) is purely {\it imaginary}. It
is well known that the Bessel function $J_0(ix)$ with
$x\in\mathbb{R}$ has no zeroes [that is, $\psi(x_{\text{m}})\neq 0$
for $x_{\text{m}}\in\mathbb{R}$]. One therefore concludes that
stationary solutions (with $\omega=m\Omega_{\text{H}}$) of the wave
field are {\it not} compatible with the mirror-like boundary
condition (\ref{Eq15}) for confining mirrors in the region
$x_{\text{m}}\ll\tau$ [see Eq. (\ref{Eq35})].

The proof presented in this section supports our previous conclusion
that, the composed Kerr-black-hole-mirror-field system is
characterized by a {\it finite} asymptotic value [see Eq.
(\ref{Eq29})] of the dimensionless ratio $x^*_{\text{m}}/\tau$,
where $x^*_{\text{m}}$ is the critical mirror radius.

\section{Summary and discussion}

In this paper, we have used analytical tools in order to study the
{\it stationary} (marginally-stable) resonances of the composed
black-hole-mirror-field system. These resonances are fundamental to
the physics of confined bosonic fields in black-hole spacetimes. In
particular, these resonances mark the onset of superradiant
instabilities in the black-hole bomb mechanism of Press and
Teukolsky \cite{PressTeu2}.

We have derived the characteristic resonance condition (\ref{Eq21})
for the marginally-stable (stationary) black-hole-mirror-field
configurations. In particular, it was shown that, for
rapidly-rotating black holes, the stationary resonances of the
system are described by the simple zeroes of the hypergeometric
function.

The characteristic resonance condition (\ref{Eq21}) determines the
discrete set of mirror radii, $\{x^{\text{stat}}_{\text{m}}(\bar
a,l,m;n)\}$, which support stationary confined field configurations
in the black-hole spacetime. One nice feature of this resonance
condition lies in the fact that it immediately reveals that the
critical (`stationary' \cite{Noteccrit}) mirror radii scale linearly
with the black-hole temperature \cite{Notelin}, see Eq.
(\ref{Eq22}). This fact implies that
$x^{\text{stat}}_{\text{m}}(\bar a,l,m;n)$ is a decreasing function
of the black-hole rotation parameter $\bar a$
--- the larger the black-hole spin, the closer to the black-hole
horizon the confining mirror can be placed.

It was shown that the stationary mirror radius,
$x^{\text{stat}}_{\text{m}}(m)$, decreases monotonically with
increasing values of the azimuthal harmonic index $m$ of the
confined field mode. In particular, our results provide compelling
evidence that the composed Kerr-black-hole-mirror-field system is
characterized by a {\it finite} asymptotic value of the critical
mirror radius: $x^{\text{stat}}_{\text{m}}(m\gg1)/\tau\simeq 0.36$
\cite{Notent}.

The physical significance of the asymptotic stationary mirror
radius, $x^*_{\text{m}}\equiv x^{\text{stat}}_{\text{m}}(\bar a\to
1,l=m\to\infty)$, lies in the fact that it is the {\it innermost}
location of the confining mirror which allows the extraction of
rotational energy from spinning black holes. This implies that, for
generic confined field configurations \cite{Noteclo}, this critical
mirror radius marks the onset of superradiant instabilities in the
composed system: composed black-hole-mirror-field systems whose
mirror radii lie in the regime $x_{\text{m}}<x^*_{\text{m}}$ are
stable (that is, all modes of the confined field decay in time),
whereas composed black-hole-mirror-field systems whose mirror radii
lie in the regime $x_{\text{m}}>x^*_{\text{m}}$ are unstable (that
is, there are confined field modes which grow exponentially over
time).

\bigskip
\noindent
{\bf ACKNOWLEDGMENTS}
\bigskip

This research is supported by the Carmel Science Foundation. I thank
Yael Oren, Arbel M. Ongo and Ayelet B. Lata for helpful discussions.


\begin{thebibliography}{99}
\bibitem{Noteun} We use natural units in which $G=c=\hbar=1$.
\bibitem{Notea} Here $r_+$ and $a$ are respectively the horizon-radius and the angular-momentum per unit mass
of the black hole, see Eq. (\ref{Eq6}) below.
\bibitem{NoteRN} An analogous extraction of Coulomb energy and electric charge may occur when
charged scalar fields in the superradiant regime $\omega<qQ/r_+$
scatter off a charged Reissner-Nordstr\"om black hole
\cite{Bekch,Dego,Hodch,Li}. Here $Q$ and $q$ are respectively the
electric charge of the Reissner-Nordstr\"om black hole and the
charge coupling constant of the field \cite{Hodch}.
\bibitem{Notelm} Here $\bar a\equiv J/M^2$ is the dimensionless angular momentum of the black hole and
$(l,m)$ are respectively the spheroidal harmonic index and the
azimuthal harmonic index of the confined field mode, see Eq.
(\ref{Eq10}) below.
\bibitem{Notelim} This fact should be contrasted with the characteristic
equation (\ref{Eq4}) which is only valid in the regime $m\ll 1/\bar
a$.
\bibitem{Noteclo} That is, confined scalar clouds which are composed of all
possible field modes (all possible values of the harmonic field
parameters $l$ and $m$).
\bibitem{Noteknt} This equation is known as the Teukolsky wave
equation.
\bibitem{Noteregu} The regularity requirements are imposed at the
two poles $\theta=0$ and $\theta=\pi$.
\bibitem{Notedeq} Below we shall see that equatorial field modes with $l=m\geq2$, the
modes we shall be interested in in this study, are characterized by
this property.
\bibitem{Notedel} One can choose $\delta>0$ without loss of generality.
\bibitem{Notelin} Note that the resonance condition (\ref{Eq21}) for the stationary
confined states is expressed in terms of the dimensionless ratio
$x^{\text{stat}}_{\text{m}}/\tau$.
\bibitem{Notem1} Here we have used the relation $-1=e^{i\pi(2n-1)}$ which implies $\ln(-1)=i\pi(2n-1)$,
where $n$ is an integer.
\bibitem{Noten0} We find from (\ref{Eq27}) that modes with $n\leq 0$ are characterized by
$x^{\text{stat}}_{\text{m}}/\tau<10$. These modes violate the
assumption (\ref{Eq23}) and thus should not be regarded as physical
ones. [It is worth emphasizing again that the analytical formula
(\ref{Eq27}) for the critical (stationary) mirror radii is valid in
the regime $\tau\ll x^{\text{stat}}_{\text{m}}\ll1$, see Eqs.
(\ref{Eq18}) and (\ref{Eq23})].
\bibitem{Notetau} It is worth emphasizing again that the mirror-like boundary
condition (\ref{Eq21}) is valid for mirror radii in the
$x_{\text{m}}\ll1$ regime [see Eq. (\ref{Eq18})]. Solving the
resonance condition (\ref{Eq21}) for small and moderate values of
the spheroidal harmonic index $l$, one finds
$x^{\text{stat}}_{\text{m}}(\bar a,m;n=1)/\tau\gtrsim 1$, see Table
\ref{Table2}. Thus, the stationary mirror radii presented in Table
\ref{Table2} are valid in the regime of rapidly-rotating
(near-extremal) black holes with $\tau\lesssim
x^{\text{stat}}_{\text{m}}(\bar a,m;n=1)\ll1$.
\bibitem{Notedel2} We are interested here in the relation
$x^{\text{stat}}_{\text{m}}=x^{\text{stat}}_{\text{m}}(\tau)$ to
leading-order in the small parameter $\tau\ll1$ (note that $\tau\ll
1$ corresponds to $\bar a\simeq 1$). We therefore take
$a\omega=m/2+O(\tau)$ and $k=m+O(\tau^2)$ in Eqs. (\ref{Eq12}),
(\ref{Eq16}), and (\ref{Eq20}). The leading-order values of the
separation constants (the angular eigenvalues) ${\delta_{lm}}$ are
given in Table \ref{Table2}.
\bibitem{Notef} In the asymptotic $m\gg1$ regime, the angular separation
constants $\delta$ are given by Eq. (\ref{Eq39}) below
\cite{Hoddel}.
\bibitem{Notecharge} Here $Q$ and $q$ are respectively the electric charge of the Reissner-Nordstr\"om black hole
and the charge coupling constant of the confined field \cite{Hodch}.
\bibitem{Noteprac} The answer to this question may one day be of practical
importance: In order to save construction materials, one would like
to build the confining mirror as close as possible to the black-hole
horizon.
\bibitem{Notene} Note that the near-horizon region (\ref{Eq35})
corresponds to $r-r_+\ll r_+-r_-$, which implies $\Delta\simeq
(r-r_+)(r_+-r_-)$.
\bibitem{Notey} Note that $y\to -\infty$ (and thus $e^{\tau y/M}\to 0$) in the near-horizon
region (\ref{Eq35}).
\bibitem{Notext} We recall that our analysis in this section is
valid in the regime $x/\tau\ll1$ [see Eq. (\ref{Eq35})]. However,
since $\lambda=O(l^2)$ [see Eq. (\ref{Eq40})], the argument
$\sqrt{\big(\lambda+{{\tau r_+}\over{2M}}\Big){{x}\over{\tau}}}$ of
the Bessel function in (\ref{Eq46}) can be large in the asymptotic
eikonal $l\gg1$ regime.
\end{thebibliography}
\end{document}